\documentclass[amsmath,amssymb,proc,secnumarabic,twocolumn]{revtex4}

\usepackage{graphicx}

\begin{document}
\title{Phonon surface mapping of graphite: disentangling quasi--degenerate phonon dispersions}


\author{A.~Gr\"{u}neis$^{1,2}$, J.~Serrano$^{3}$
A.~Bosak$^{4}$, M.~Lazzeri$^{5}$,
S.L.~Molodtsov$^6$, L.~Wirtz$^7$, C.~Attaccalite$^8$, M.~Krisch$^4$, A.~Rubio$^8$, F.~Mauri$^{5}$,
T.~Pichler$^1$}

\affiliation{$^{1}$Faculty of Physics, University of Vienna,
Boltzmanngasse 5, A-1090 Vienna,
Austria}\affiliation{$^{2}$IFW--Dresden, P.O. Box 270116, D-01171
Dresden, Germany}\affiliation{$^{3}$ICREA-Departament de Fisica
Aplicada, EPSC, Universitat Politecnica de Catalunya, Av. Canal
Olimpic 15, 08860 Castelldefels, Spain}\affiliation{$^{4}$European
Synchrotron Radiation Facility, BP 220, F-38043 Grenoble Cedex,
France } \affiliation{$^{5}$IMPMC, Universit\'es Paris 6 et 7,
CNRS, IPGP, 140 rue de Lourmel, 75015 Paris, France}
\affiliation{$^{6}$Institut f\"ur Festk\"orperphysik, TU Dresden,
Mommsenstrasse 13, D-01069 Dresden, Germany
}\affiliation{$^{7}$IEMN - Dept. ISEN, CNRS-UMR 8520, Villeneuve
d'Ascq, France} \affiliation{$^{8}$Nano-Bio Spectroscopy Group and
European Theoretical Spectroscopy Facility (ETSF), Departamento de
Fisica de Materiales, Unidad de Materiales Centro Mixto
CSIC-UPV/EHU, Universidad del Pais Vasco, Avd. Tolosa 72, E-20018
Donostia, Spain}

\begin{abstract}
The two-dimensional mapping of the phonon dispersions around the $K$ point of graphite by
inelastic x-ray scattering is provided. The present work resolves
the longstanding issue related to the correct assignment of transverse and
longitudinal phonon branches at $K$. We observe an
almost degeneracy of the three TO, LA and LO derived phonon branches and a strong
phonon trigonal warping. Correlation effects renormalize the Kohn anomaly
of the TO mode, which exhibits a trigonal warping effect opposite to that
of the electronic band structure. We determined the electron--phonon coupling constant
to be 166$\rm(eV/\AA)^2$ in excellent agreement to $GW$ calculations. These results
are fundamental for understanding angle-resolved photoemission, double--resonance Raman and transport
measurements of graphene based systems.
\end{abstract}
\maketitle

The lattice dynamics of most ``standard'' materials are well
known nowadays and can be routinely described by {\it ab-initio}
methods based on density--functional perturbation theory \cite{DFPT}.
The phonon dispersion relations of graphite and graphene have already
been determined by inelastic neutron scattering~\cite{D105},
inelastic x-ray scattering (IXS)~\cite{maultzsch04,mohr07-ixs}, double--resonance Raman
scattering~\cite{c887,q953} and electron energy loss
spectroscopy~\cite{aizawa90}. After the first {\it ab-initio} calculations \cite{kresse95,pavone96}
had been corroborated by IXS~\cite{maultzsch04,mohr07-ixs,bosak07-graphit}, the story for
graphite seemed closed~\cite{ludger04-phonon}. However, the phonon
dispersions of graphite and graphene - a single layer of graphite
- have continued to present surprises in recent years:
(i) two Kohn anomalies~\cite{kohn62-anomaly} at the Brillouin zone (BZ) center
($\Gamma$) and corner ($K$) were predicted theoretically~\cite{ando06-tophonon,piscanec04-kohn} and (ii) for charged
graphene/graphite, the (electronic) non-adiabatic effects at $\Gamma$ were
predicted theoretically~\cite{lazerri06-kohn} and confirmed
experimentally by Raman scattering~\cite{mauri07-bo,kim07-elphon}.
Very similar observations have been made for carbon
nanotubes~\cite{mauri07-kohncarbon,das07-transport,farhat07-kohn,avouris07-kohn,sasaki08-kohn}.
The precise understanding of the Kohn anomaly is highly significant for
electron--phonon coupling (EPC) and thus of prime importance for the
quantitative description of superconductivity in graphite
intercalation
compounds~\cite{hannay65-gicsuperconductivity,mauri-cac6}, electronic transport in the high bias regime, and for the (double)
resonant Raman scattering~\cite{thomsen00d,c887}. The latter is
now commonly used to distinguish single-layer graphene from
double- and multi-layer graphene \cite{ferrari07-graphene,Graf07}.

Qualitatively, the Kohn anomaly of the transversal optical (TO) branch at the $K$ point is well described by
density functional theory (DFT)~\cite{piscanec04-kohn}. However,
its magnitude is severely underestimated
by DFT~\cite{basko08-elph,lazzeri08}. Electron-electron correlation leads to an enhancement of the EPC
at $K$. This can be quantitatively calculated~\cite{lazzeri08} with the $GW$--approximation
which also gives very good results for the electronic band-structure of graphite (see, e.g., Ref.~\cite{alex06-correlation}).
Although the EPC at $\Gamma$ can be derived from the IXS and Raman linewidth
measurements~\cite{lazzeri06-linewidth} the experimental
determination of the EPC close to the $K$
point~\cite{lazzeri08} is still missing. The experiments needed to unravel the details of the phonon dispersions
are extremely challenging because three phonon branches are overlapping in a small energy window
and the phonon energies have a the strong dependence on doping. Thus, in order to overcome
this problem there is an urgent need to map the full two--dimensional phonon dispersion relations around $K$ point in a similar
manner as done routinely for electron energy dispersions by angle--resolved photoemission spectroscopy (ARPES).

In this Letter we present a mapping of the detailed two--dimensional phonon
dispersion relation around the $K$ point of graphite
single crystals measured by IXS. The phonon dispersion perpendicular to the graphene layers in
graphite is practically zero for the phonon modes around
$K$~\cite{marzari05,ludger04-phonon} and thus we can use graphite
to test the phonons of graphene. We prove that at
$K$, the two branches with LA and LO character and the branch with TO character are
almost triply degenerate due to the Kohn anomaly which brings down the TO
branch in energy. The phonon mode assignment is
carried out by explicit comparison of the measured and calculated
IXS intensities in the two--dimensional map and used to disentangle overlapping phonon modes.
Using the present unambigous assignment of the phonon branches and previously measured
electron dispersions around $K$ we determine the value of the EPC entirely from
experiments and find quantitative agreement to $GW$ calculations.

\begin{figure}
\hspace{-0.5cm}\includegraphics[width=9cm]{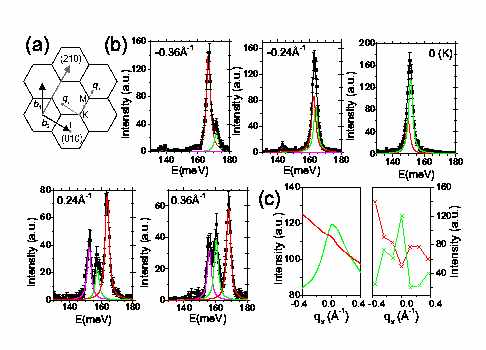}
\caption{(a) Scattering geometry for the IXS experiments: spectra
are collected along a line starting from the (010) BZ along the (210) [i.e. $\Gamma KM$] direction.
(b) IXS spectra of graphite single crystals along the $\Gamma KM$
line. The three phonon modes are fit by Lorentzians. The magenta
and green Lorentzians denote the LA and LO modes. The TO mode is
denoted by a red Lorentzian. (c) The calculated (left panel) and
measured (right panel) IXS intensites for LO (green) and TO (red)
modes.\label{fig:ixs1}}
\end{figure}

The experiments were performed on beamline ID28 at the ESRF,
utilizing the silicon (999) setup which yields monochromatized
synchrotron light of 17794~eV, providing a total energy resolution
of $\sim$3.3~meV~\cite{krisch03-id28}. The beam size at the sample
position was 60~$\times$~30~$\mu m^2$. As the spectrometer is
equipped with nine analyzers, nine IXS spectra were recorded
simultaneously, thus allowing an efficient coverage of the
relevant portion of the phonon dispersion branches. Single
crystalline graphite samples with about 1~cm diameter and 100~$\mu
m$ thickness were mounted on a goniometer. In
Fig.~\ref{fig:ixs1}(a) the scattering geometry and the coordinate
system used throughout this work is shown.  We collected IXS
spectra along the line starting from the (010) reciprocal lattice
point in the (210) direction. We have purposely chosen this line
since calculations of the IXS intensities reveal the highest
intensity for the TO mode in the (010) BZ.

\begin{figure*}[t]
\includegraphics[width=14cm]{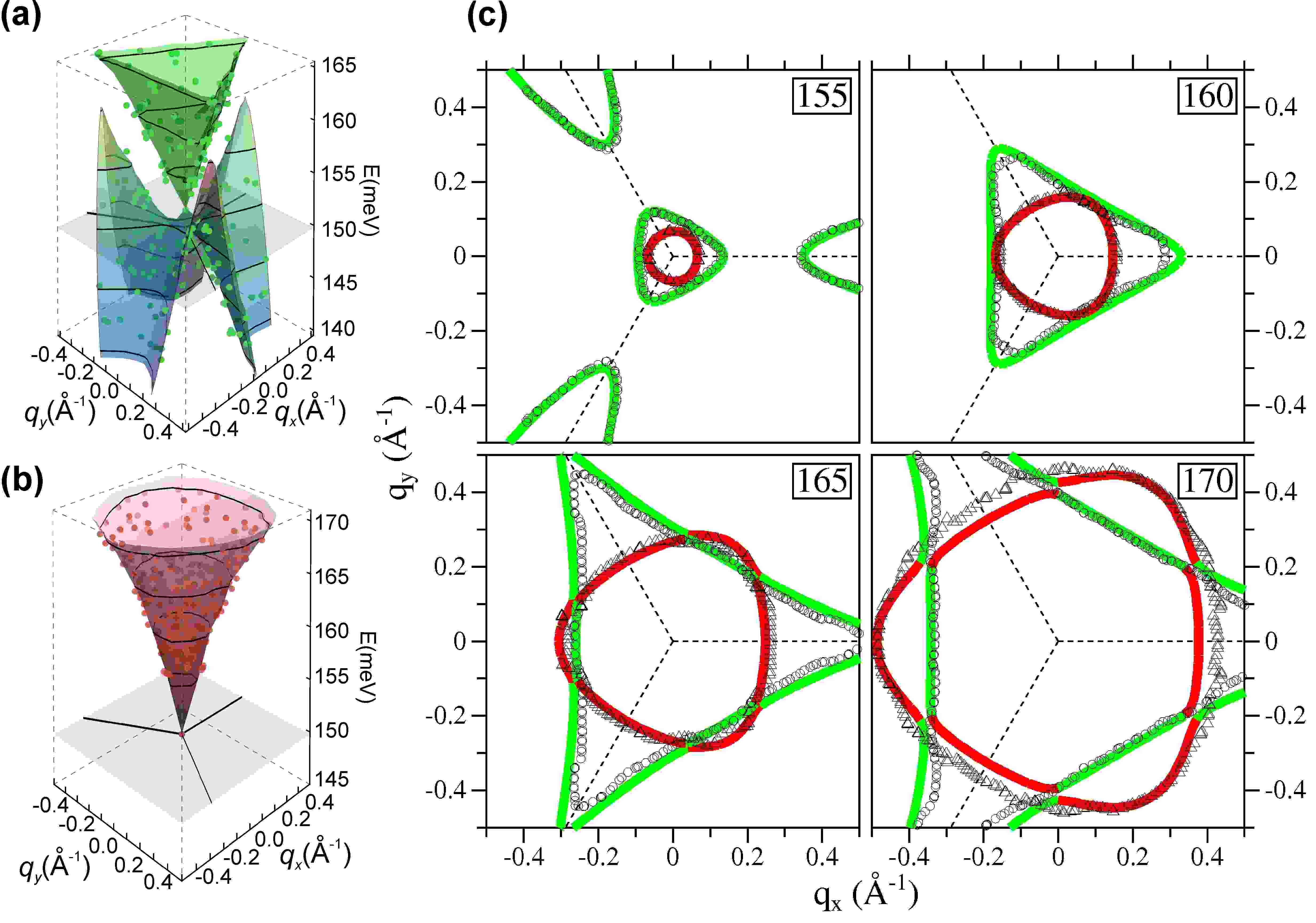}
\caption{ The interpolated experimental two--dimensional phonon
dispersion relation for (a) the longitudinal optic and accoustic
phonon branches and (b) the transverse optical phonon branch that
has a Kohn anomaly. Dots denote the experimental data taken from the maxima
of Lorentzian fits [see Fig.~1(b)]. (c) Equi--energy contours from
the interpolated phonon surfaces for the three highest phonon
branches around $K$ for energies 155~meV--170~meV. $\circ{}$
and $\bigtriangleup$ denote an interpolation to the raw
experimental data in (a) and (b). Lines are $GW$ calculations.
The present color code follows the diabatic
notation: branches with the same color have similar phonon
patterns.\label{fig:equi}}
\end{figure*}
The phonon frequencies are the eigenvalues of $\sqrt{{\cal D}_{\bf
q}/M}$, where ${\cal D}_{\bf q}$ is the dynamical matrix at the
wavevector {\bf q} and $M$ is the carbon mass. The DFT dynamical
matrix ${\cal D}^{DFT}_{\bf q}$ is computed using linear
response~\cite{DFPT}, with the exchange-correlation of
Ref.~\cite{PBE} and other computational details as in
Ref.~\cite{piscanec04-kohn}. The Kohn anomaly of the TO branch near $K$ is
entirely determined~\cite{piscanec04-kohn,lazzeri08} by the
contribution (to ${\cal D}_{\bf q}$) of the phonon self-energy
between $\pi$ bands $P_{\bf q}$ (defined in Eq.1 of
Ref.~\cite{lazzeri08}). In particular, the TO branch corresponds
to the eigenvector $|p_{\bf q}\rangle$ of $P_{\bf q}$ with the
eigenvalue $p_{\bf q}$ that is largest in modulo. $p_{\bf q}$ is
proportional to the square of the EPC between the $\pi$ bands near
the Fermi level. In Ref.~\cite{lazzeri08} it was shown that DFT
underestimates $|p_{\bf q}|$ and that $|p_{\bf q}|$ from  the more
accurate $GW$ calculations is larger by a factor $r^{GW}=1.61$.
Here, following Ref.~\cite{lazzeri08}, we compute the $GW$
dynamical matrices as
\[
{\cal D}^{GW}_{\bf q}={\cal D}^{DFT}_{\bf q}+ |p_{\bf
q}\rangle\langle p_{\bf q}| \{(r^{GW}-1)p_{\bf q}+\Delta\},
\]
where $\Delta$ is a constant independent of {\bf q}. $r^{GW}$
determines the slope of the TO branch near $K$. $\Delta$ shifts
rigidly the TO branch with respect to the other branches. $\Delta$
cannot be computed easily $ab-initio$ (see footnote 24 of
Ref.~\cite{lazzeri08}) and is fitted to reproduce the present
measurements. Furthermore DFT slightly overestimates (in a
systematic manner) the phonon frequencies of the branches not
affected by the Kohn anomaly, thus all the frequencies are scaled by -1.3 \%
to simplify the comparison with measurements.

Selected raw IXS spectra along the $\Gamma KM$ direction are shown
in Fig.~\ref{fig:ixs1}(b). These spectra
contain the contributions of three phonon modes. A lineshape
analysis using Lorentzians yields the position and intensity of
each mode which are assigned to the LA, LO and TO branches. Note
that this characterization of phonon eigenvectors is only strictly
valid for very small $k$ away from $\Gamma$ and we identify the
phonon branches in the whole BZ according to their
characterization close to $\Gamma$~\cite{marzari07}. From symmetry
considerations LO and LA modes are degenerate at the $K$ point.
Our results clearly show that LO/LA and TO are almost degenerate
at and close to the $K$ point. This is in striking contrast to
previously determined phonon
dispersions\cite{maultzsch04,mohr07-ixs} and can not be explained
by the standard DFT phonon calculations~\cite{maultzsch04}. Our
mode assignment is based on a comparison of the measured and
calculated IXS intensities as shown in Fig.~\ref{fig:ixs1}(c) and
further supported by a comparison to the $GW$ calculated phonon
frequencies as shown later. It is clear that the calculated and
experimental IXS intensities of LO and TO show the same pattern
and thus we can safely assign the TO mode as the one with higher
intensity along the considered path in the (010) BZ.

From the maxima of the Lorentzians a set of $\sim$100 data points
is obtained in the $q_xq_y$ plane in a region of
$\pm$0.6~$\rm\AA^{-1}$ around the $K$ point, which is further
extended by the application of symmetry operations. In
Fig.~\ref{fig:equi}(a,b) the raw experimental data points around
the $K$ point are shown. From these data points the
three--dimensional surfaces of the phonon dispersion relations are
interpolated using spline functions. The interpolated phonon maps
are used for further data analysis. The phonon trigonal warping
effect for the LA,LO and TO modes around the $K$ point is
illustrated from the phonon equi--energy contours shown in
Fig.~\ref{fig:equi}(c) along with the $GW$ calculations. Notably,
the experiments and calculations are in excellent agreement to
each other which further supports the mode assignment performed
above. Interestingly, the phonon trigonal warping effect of the TO
mode is opposite to the electronic trigonal warping effect, i.e.
the slope of the TO phonon branch is higher in $KM$ direction,
whereas the electronic bands have a higher slope in $K\Gamma$
direction~\cite{alex06-correlation}. For the LA and LO modes, the
phonon trigonal warping has the same angular dependence as for the
electronic bands. The fact that the equi--energy contour of the TO
mode is almost circular, is also in agreement with the angular
dependence of the kink due to EPC in the quasiparticle dispersion
measured by ARPES, where no change of the kink
position, i.e. the energy of the coupling
phonon~\cite{alex08-kc8} was observed. Furthermore, the observed contour for
the TO mode can explain why the Raman linewidth of the G' peak
depends weakly on the exciting laser energy. The trigonal warping of electrons and
the TO phonons is opposite, canceling the effects of
enlarging/shrinking of the Raman linewidths. Furthermore it is clear that the LO and TO branches
cross along the high-symmetry $K\Gamma$ direction
but undergo anti-crossing along all the other directions.
\begin{figure}
\includegraphics[width=8cm]{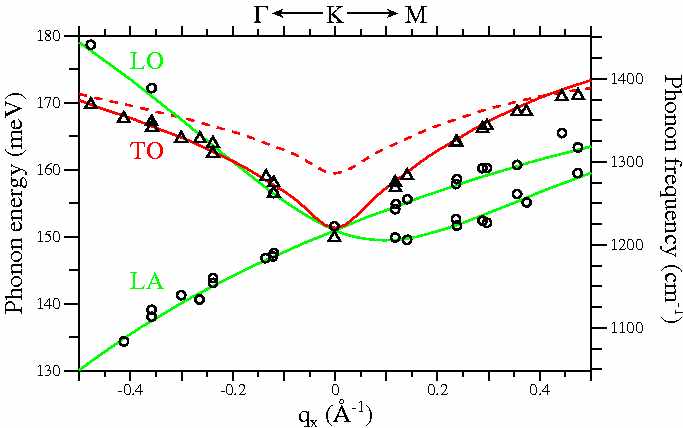}
\caption{Phonon dispersion along  the $\Gamma KM$ direction.
Points are measurements extracted from the maxima of the
Lorentzians from Fig.~1(b). Full lines are $GW$ calculations. The
dashed line is the TO branch from DFT calculations. The color code
is the same as in Fig.~\ref{fig:equi}. \label{fig:ixs2}}
\end{figure}
In a final Fig.~\ref{fig:ixs2} we show the phonon dispersions along the high--symmetry
lines and compare it to previous DFT calculations and our $GW$ results.
It can be observed that (1) the three branches are almost touching
each other at $K$ (2) the LO and TO branches cross each other in
$K\Gamma$ direction and (3) the proper inclusion of the
electron-electron correlation (through the $GW$ approach) in the
calculation of the EPC is crucial to obtain a good agreement with
the experiment. Indeed $GW$ calculations are in almost perfect
agreement with measurements while DFT calculations severely
underestimate the phonon slope of the TO branch and thus the EPC
at $K$ and overestimate the phonon energy. For precise description of the experimental phonon
dispersion relation, we provide simple fit formulas for phonon
dispersions of the TO and LO branches. The TO phonon dispersion is
given by
\begin{equation}
E_{TO}(q,\theta)=2.56 \cos (3 \theta ) q^2-62.75 q^2+73.07 q+149.8
\label{eq:to}
\end{equation}
and the LO phonon dispersion is best fit by
\begin{equation}
\begin{array}{lll}
 E_{LO}(q,\theta)=&-10.25 q \cos ^4(3 \theta )-5.32q \cos ^3(3 \theta )+ &   \\ & + 5.47q \cos ^2(3 \theta )+ 26.44 q^2+37.01
q- &  \\
& -25.19 q^2 \cos (3 \theta )+151.5. &
\end{array}
\end{equation}
Here $E_{TO}(q)$ and $E_{LO}(q)$ denote the phonon energies in meV
of TO and LO branch, respectively, $q$ is the phonon wavevector
measured in units $\rm\AA^{\rm -1}$ from $K$ point [see Fig. 1(a)]
and $\theta$ the angle away from the $KM$ direction. From these
fits we obtain the slope of the TO branch at $K$,
$S_{TO}^{K}=73.07$meV$\rm\AA$, from which we can derive an
experimental value for the square of the EPC constant between
$\pi$ bands at $K$, $\langle
D^2_K\rangle_F$~\cite{piscanec04-kohn}. Indeed, (see Eq. (10)
of~\cite{piscanec04-kohn}),
\begin{equation}
\langle D^2_K\rangle _F=\frac{8S_{TO}^{K}M\hbar
\omega_{TO}^Kv_f}{\sqrt{3}\hbar a_0^2},
\label{eq:d}\end{equation} where $\langle D^2_K\rangle _F$ is
defined in~\cite{lazzeri08}, and all the parameters on the right
hand side were experimentally determined in this work or
previously by ARPES~\cite{alex06-correlation}. In particular,
$\hbar\omega_{TO}^K=149.8$meV is the the TO phonon energy at $K$,
$v_F=1.05\times
10^6$ms$^{-1}$~\cite{alex06-correlation,alex07-tbgw} the Fermi
velocity, $M$ the carbon atom mass and $a_0=2.46\rm\AA$ the in-plane
lattice constant. Using Eq.(\ref{eq:d}), we obtain $\langle
D^2_K\rangle_F=166$~$\rm(eV/\AA)^2$ which is in excellent
agreement to the theoretical $GW$ value of $\langle
D^2_K\rangle_F=164\rm(eV/\AA)^2$~\cite{lazzeri08} and is to be
compared to much lower DFT value of $\langle
D^2_K\rangle_F=92\rm(eV/\AA)^2$~\cite{lazzeri08}. Finally,
concerning the temperature dependence of the energy ordering and
the LO-TO difference at $K$, we performed experiments at three
different temperatures (300K, 150K and 15K) and find no change in
the ordering and a slight increase in the energy difference at $K$
with decreasing temperature (2.2~meV instead of 0.7~meV).

In summary we have disentangled overlapping phonon branches
and mapped the 2D phonon surfaces at the $K$ point of graphite using a
combination of IXS and $ab-initio$ calculations. Excellent
agreement for the phonon energies was found once many--body effects at the $GW$ level are
considered. The three highest branches around $K$ point are almost
triply degenerate, in stark contrast to previous measurements and
DFT calculations~\cite{maultzsch04,mohr07-ixs}. Furthermore the Kohn anomaly
of the TO phonon branch at $K$ was directly measured by IXS and
the EPC was determined entirely from experiments. An opposite
trigonal warping of the TO phonon branch as compared to the
electrons and the LO phonon branch was found. This is important
for understanding the angular dependence of the kink in the
quasiparticle dispersion as measured by ARPES in graphite
intercalation compounds and graphene and the linewidth of double--resonant Raman scattering.

A.G. acknowledges an APART fellowship from the Austrian Academy of
Sciences and a Marie Curie Individual Fellowship (COMTRANS)
from the EU. J.S. acknowledges support from Spanish MICINN
(grants MAT2007-60087 and ENE2008-04373) and Generalitat de Catalunya (2005SGR00535). AR and LW are funded by
Spanish MEC (FIS2007-65702-C02-01), "Grupos Consolidados UPV/EHU
del Gobierno Vasco" (IT-319-07) and  EC-I3 ETSF project (Contract
211956). Calculations were done at IDRIS and Mare Nostrum
 "Red Espanola de Supercomputacion".

\end{document}